\documentclass[floats, aps, pre, showpacs, nofootinbib, twocolumn]{revtex4-1}
\usepackage{bm,amsfonts, mathtools}
\usepackage{graphicx,color, wasysym}
\usepackage{textcomp}
\usepackage{amsmath,amssymb,latexsym,epsfig}

\def\ulamek#1#2{\mbox{\normalfont$\frac{#1}{#2}$}}

\begin{document}

\makeatletter

\title{The Higher-Order Heat-Type Equations via signed L\'{e}vy stable and generalized Airy functions}

\author{K.~G\'{o}rska}
\email{katarzyna.gorska@ifj.edu.pl}
\affiliation{H. Niewodnicza\'{n}ski Institute of Nuclear Physics, Polish Academy of Sciences, ul.Eljasza-Radzikowskiego 152, 
PL-31342 Krak\'{o}w, Poland}

\author{A.~Horzela}
\email{andrzej.horzela@ifj.edu.pl}
\affiliation{H. Niewodnicza\'{n}ski Institute of Nuclear Physics, Polish Academy of Sciences, ul.Eljasza-Radzikowskiego 152, 
PL-31342 Krak\'{o}w, Poland}

\author{K.~A.~Penson}
\email{penson@lptl.jussieu.fr}
\affiliation{Laboratoire de Physique Th\'eorique de la Mati\`{e}re Condens\'{e}e,\\
Universit\'e Pierre et Marie Curie, CNRS UMR 7600\\
Tour 13 - 5i\`{e}me \'et., B.C. 121, 4 pl. Jussieu, F-75252 Paris Cedex 05, France\vspace{2mm}}

\author{G. Dattoli}
\email{dattoli@frascati.enea.it}
\affiliation{ENEA - Centro Ricerche Frascati, via E. Fermi, 45, IT-00044 Frascati (Roma), Italy}

\pacs{05.10.Gg, 05.40.-a, 02.30.Uu}

\begin{abstract}
We study the higher-order heat-type equation with first time and $M$-th spatial partial derivatives, $M = 2, 3, \ldots$. We demonstrate that its exact solutions for $M$ even can be constructed with the help of signed L\'{e}vy stable functions. For $M$ odd the same role is played by a special generalization of Airy Ai function that we introduce and study. This permits one to generate the exact and explicit heat kernels pertaining to these equations. We examine analytically and graphically the spacial and temporary evolution of particular solutions for simple initial conditions. 

\end{abstract}

\maketitle


\section{Introduction}\label{Intro}

The Brownian motion governed by the conventional heat equation \cite{DVWidder75} has several generalizations. Some of them are related to the Markov processes described by the second order partial differential equation \cite{FMCholewinski68, DTHaimo71, WAWoyczynski05}. The other ones are connected with the so-called one- and two-sided L\'{e}vy stable distribution \cite{KAPenson10, KGorska11a, APiryatinska05}. The last ones are governed by the so-called higher-order heat-type equations (HOHTE)
\begin{equation}\label{E1}
\frac{\partial}{\partial t} F_{M}(x, t) = \kappa_{M} \frac{\partial^{M}}{\partial x^{M}} F_{M}(x, t), 
\end{equation}
which for integer $M > 2$, are associated with a pseudo-Markov processes (signed processes). We choose to normalize $F_{M}(x, t)$ according to $\int_{-\infty}^{\infty} F_{M}(x, t) dx = 1$. The pseudo Markov processes were introduced in the sixties of the last century and have been studied in many papers starting from \cite{VYuKrylov60, YuLDaletsky69, KJHochberg78}. Eq. (\ref{E1}) for $M=4$ is called the biharmonic heat equation \cite{APSSelvadurai00, FGazzola08}.

HOHTE of order 3 or 4 (or higher) have been intensively investigated by several authors and display many interesting features  \cite{VYuKrylov60, KJHochberg78, KJHochberg94, XLi93, ALachal03} like \textit{e.g.} oscillating nature, connection to the arc-sine law and its counterpart, the central limit theorem and so on. HOHTE arise in physical phenomena, \textit{e.g.}, in the fluctuation phenomena in chemical reactions \cite{CWGardiner77, CWGardiner79} or as a new method for imaginary smoothing based on the biharmonic heat equation \cite{MLysaker03, MRHaijaboli09}. The biharmonic heat equation is used for describing the diffusion on the unit circle \cite{JBGreer06}. For fixed values of integer $M > 2$ HOHTE can be considered as the composition of Brownian motions or stable processes with Brownian motions \cite{TFunaki79, KJHochberg96}.

In Eq. (\ref{E1}) the constants $\kappa_{M}$ are subject of constraints. Following \cite{ALachal03, KJHochberg94} we choose
\begin{equation}\label{E2}
\kappa_{M} = \left\{\begin{array}{c c}
(-1)^{M/2 + 1}, & M = 2, 4, \ldots, \\[0.6\baselineskip] 
\pm (-1)^{[M/2] + 1}, & M = 3, 5, \ldots.
\end{array}\right.
\end{equation}
The symbol $[n]$ denotes the integer part of $n$. This choice of Eq. (\ref{E2}) for $\kappa_{M}$ warrants, as we will see in Sec. \ref{IT}, the possibility of getting the solution of Eq. (\ref{E1}) via appropriate integral transform. Moreover, such a choice of coefficients $\kappa_{M}$ guarantees a holding of the classical arc-sine law for even $M$ \cite{VYuKrylov60} and the counterpart to that law in the case of odd $M$ \cite{KJHochberg94}.

From mathematical point of view Eq. (\ref{E1}), with the initial condition $F_{M}(x, 0) = f(x)$, is the Cauchy problem. Its formal solution is obtained by using the extension of the evolution operator formalism, introduced by Schr\"{o}dinger, that gives
\begin{equation}\label{E3}
F_{M}(x, t) = \hat{U}_{M}(t) f(x), \quad \hat{U}_{M}(t) = \exp\left(\kappa_{M} t \frac{\partial^{M}}{\partial x^{M}} \right)
\end{equation}
with $f(x)$ being an infinitely differentiable function, or an appropriate limit of a sequence of such functions, see below.

Below we shall employ Eqs. (\ref{E3}) to solve Eq. (\ref{E1}) for given initial conditions. For instance, for $f(x) = x^{n}$ ($n\in\mathbb{N}$) a solution of HOHTE can be expressed with the Hermite-Kamp\'{e} de F\'{e}ri\'{e}t polynomials $H^{(M)}_{n}(x, y)$ \cite{PAppell26, GDattoli07,GDattoli} as
\begin{equation}\label{E4}
F^{(n)}_{M}(x, t) = H_{n}^{(M)}(x, \kappa_{M}t) = n! \sum_{r=0}^{[n/M]} \frac{(\kappa_{M} t)^{r} x^{n-Mr}}{r! (n-Mr)!}.
\end{equation}
For $M=2$, $H^{(2)}_{n}(x, y)$ are known as the heat polynomials \cite{PCRosenbloom59, DVWidder75}. Any initial function in the form of power series, \textit{i.e.} $f(x) = \sum_{n=0}^{\infty}a_{n} x^{n}$, allows $F_{M}(x, t)$ to be represented as the following expansion
\begin{equation}\label{E5}
F_{M}(x, t) = \sum_{n=0}^{\infty} a_{n}\, H_{n}^{(M)}(x, \kappa_{M}t).
\end{equation}
In the general case the above representation of $F_{M}(x, t)$ can lead to the divergent series even for the well-defined initial condition $f(x)$. Nevertheless, Eq. (\ref{E5}) gives the correct asymptotic expansion of $F_{M}(x, t)$. The formal solution (\ref{E5}) is not effective because it converges for short times only, \textit{e.g.}, for $M = 2$ and $f(x)~=~\exp[-x^{2}/(2\sigma^{2})]/(\sqrt{2\pi} \sigma)$ the convergence is limited to $t < \sigma^{2}/(4 k)$ \cite{DVWidder63, KGorska12b}. (However, observe that such initial conditions as those before Eq. \eqref{E5} are not integrable.) The correct long time behavior of a solution of the heat equation is provided by the Gauss-Weierstrass transform for $M=2$ \cite{DVWidder75}. Therefore we look for an analogous transform for $M =3, 4, \ldots$\,\,.

The main purpose of the paper is to find the new type of integral transform which will be furnishing the long time-behavior of the formal solution (\ref{E3}) for integer $M > 2$. The paper is organized as follows. In Sec. \ref{IT} we will develop the operational methods initiated in \cite{DBabusci11, GDattoli09, DVWidder79} for generalizing the Gauss-Weierstrass transform. We will show that such a new type of integral transform is well-defined as its kernel is converging. Next, we will find the integral representation of the evolution operator. Secs. \ref{Levy} and \ref{Airy} are devoted to finding the exact and explicit forms of the integral kernels of obtained transform. The asymptotic expansion for small and large values of argument as well as the Mellin transform of the integral kernel will be considered. In Sec. \ref{Levy} we study the L\'{e}vy signed functions associated with the even values of $M$. The case of odd $M$ and the generalized Airy function is investigated in Sec. \ref{Airy}. The specific examples of solutions of HOHTE are presented in Sec. \ref{SExp}. In this Section we also derive some new formulas of Glaisher-type. We conclude the paper in Sec. \ref{Conc}.

\section{Integral transform}\label{IT}

Following the example for $M=3$ developed in \cite{DVWidder79}, we consider the integral
\begin{eqnarray}\label{E6}
p_{M}(x, t) = \frac{1}{2\pi i} \int_{c - i\infty}^{c + i\infty} e^{x s + \kappa_{M} t s^{M}} ds
\end{eqnarray}
with $s = c + i \tau$ and $c$, $\tau \in \mathbb{R}$. The quantity $p_{M}(x, t)$ of Eq. (\ref{E6}) will play the role of higher-order heat kernel needed to represent the solutions of Eq. (\ref{E1}) via appropriate integral transform. If $s = |s| e^{i \varphi}$, $|s| = \sqrt{c^{2} + \tau^{2}}$ and $\varphi = \arctan(\ulamek{\tau}{c})$, then for $t > 0$, the following relations hold 
\begin{align}\label{E7}
\left\vert p_{M}(x, t)\right\vert  &\leq \int_{-\infty}^{\infty} e^{\Re\left(x s + \kappa_{M} t s^{M}\right)} d\tau \nonumber\\
&= \int_{-\infty}^{\infty} e^{x c + \kappa_{M} t |s|^{M} \cos(M\varphi)} d\tau.
\end{align}
For large $\tau$ we have $|s| \approx \tau$, $\varphi \approx \ulamek{\pi}{2} - \ulamek{c}{\tau}$ and $\cos(M\varphi) \approx \cos(M\ulamek{\pi}{2}) + M \ulamek{c}{\tau}\sin(M\ulamek{\pi}{2})$, where for $M = 2m$, $\cos(2m\varphi) \approx (-1)^{m}$ and for $M=2m+1$, $\cos[(2m+1)\varphi] \approx (2m+1)(-1)^{m} \ulamek{c}{\tau}$. For large $\tau$, that gives Eq. (\ref{E7}) in the form
\begin{equation}\label{E8}
\left\vert p_{M}(x, t)\right\vert \leq \int_{-\infty}^{\infty} e^{x c + \kappa_{2m} (-1)^{m} t \tau^{2m}} d\tau, 
\end{equation}
where $M=2m$, and 
\begin{equation}\label{E9}
\left\vert p_{M}(x, t)\right\vert \leq \int_{-\infty}^{\infty} e^{x c + (2m+1) \kappa_{2m+1} (-1)^{m} c t \tau^{2m}} d\tau
\end{equation}
with $M=2m+1$. The integral of Eq. (\ref{E8}) converges only for $\kappa_{2m} = (-1)^{m+1}$, whereas Eq. (\ref{E9})  converges for two values of $\kappa_{M}$, \textit{i.e.} for $c > 0$ we have $\kappa_{2m+1} = (-1)^{m+1}$ and for $c < 0$, $\kappa_{2m+1} = (-1)^{m}$. That substantiates the conditions specified in Eq. (\ref{E2}).

Moreover, applying to Eq. (\ref{E6}) the Cauchy's theorem with integration over the rectangle $s = \pm i R$, $c \pm i R$, it can be shown that in Eq. (\ref{E6}) we can take $c = 0$. As $|R|$ tends to infinity, the integrals over horizontal sides approach zero. It boils down to two inequalities:
\begin{equation}\label{e3}
\Big\vert\int_{\pm iR}^{c\pm iR} e^{x \tilde{s} + \kappa_{2m} t \tilde{s}^{2m}} ds\Big\vert \leq e^{- t R^{2m}} \int_{0}^{c} e^{x\sigma} d\sigma,
\end{equation}
for $M=2m$ and $\tilde{s} = \sigma \pm iR$, and 
\begin{equation}\label{e4}
\Big\vert\int_{\pm iR}^{c\pm iR} e^{x \tilde{s} + \kappa_{2m+1} t \tilde{s}^{2m+1}} ds\Big\vert \leq \int_{0}^{c} e^{x\sigma \mp (2m+1) t \sigma R^{2m}} d\sigma,
\end{equation}
for $M=2m+1$, that vanish in the limit of $|R|\to \infty$. Consequently the integral Eq. (\ref{E6}) converges (for $M = 2m$ absolutely) when $c=0$. Thus for $t > 0$ 
\begin{equation}\label{E10}
p_{M}(x, t) = \frac{1}{2\pi} \int_{-\infty}^{\infty} e^{i x \tau + \kappa_{M} (i \tau)^{M} t} d\tau,
\end{equation}
which, after substitution of $\kappa_{M}$ from Eq. (\ref{E2}) and changing the variable $y = \tau t^{1/M}$, can be expressed in the form
\begin{eqnarray}\label{E11}
p_{M}(x, t) &=& \frac{1}{t^{1/M}} g_{M}\left(\frac{x}{t^{1/M}}\right), \\[0.4\baselineskip]
g_{M}(u) &=& \Re\left[\frac{1}{\pi} \int_{0}^{\infty} e^{i u y+ \kappa_{M} (i y)^{M}} d y \right]. \label{E12}
\end{eqnarray}
It is clear that $p_{M}(x, 1) = g_{M}(x)$. (In \cite{KGorska12} the integral kernels of the form of Eq. (\ref{E11}) have been already treated, however under the restriction $M = \alpha < 1$.) We point out that Eqs. (\ref{E10}), (\ref{E11}) and (\ref{E12}) are defining in the alternative way the formal solution of Eq.~(\ref{E1}) with $f(x) = \delta(x)$ \cite{KJHochberg94, ALachal03}, because in the limit of $t=0^{+}$, $p_{M}(x, t)=\delta(x)$, where $\delta(x)$ is Dirac delta function. For $M=2m$ Eq. (\ref{E10}) can be rewritten in the form
\begin{equation}\label{E13}
g_{2m}(u) = \frac{1}{\pi} \int_{0}^{\infty} \cos(u y) e^{-y^{2m}} dy,
\end{equation}
where $g_{2m}(u) = g_{2m}(-u)$ is the so-called symmetric L\'{e}vy stable signed function \cite{KGorska11a, TMGaroni02, TGaroni03}. Note that $g_{2m}(u)$ of Eq. (\ref{E13}) here, are denoted by $g(2m, 0, u)$ in Ref. \cite{KGorska11a}. It should be stressed that the functions $g^{\pm}_{2m+1}(u)$ do \textit{not} belong to the family of L\'{e}vy stable functions considered in Ref. \cite{KGorska11a}. For $M=2m+1$
\begin{equation}\label{E14}
g^{\pm}_{2m+1}(u) = \frac{1}{\pi} \int_{0}^{\infty} \cos(u y \mp y^{2m +1}) dy
\end{equation}
defines the generalization of the Airy function in the sense of \cite{RCYChin78, DBabusci11, DBabusci}. The functions $g^{\pm}_{2m+1}(u)$ are not anymore even functions and Eq. (\ref{E14}) implies $g^{+}_{2m+1}(-u) = g^{-}_{2m+1}(u)$. Without loss of the generality, in this paper we will be studying the case of $g^{+}_{2m+1}(u)$. We shall adopt the notation of \cite{DBabusci11} and henceforth set $g^{+}_{2m+1}(x) \equiv Ai^{(2m+1)}(x)$. The notation with $\pm$ will be used when necessary. The explicit and exact form of $g_{M}(u)$ will be derived in the next two Sections.

It seems that obtaining $p_{M}(x, t)$ for arbitrary integer $M \geq 2$ constituted a true challenge: Until recently only for a limited number of values of $M$, \textit{i.e.} $M=2$ \cite[see formula 2.3.15.11 on p. 344]{APPrudnikov-1}, $M=3$ \cite{DVWidder79, DBabusci11}, $M=4$ \cite{KGorska13}, and $M=6$ \cite{KGorska13} the explicit forms of $p_{M}(x, t)$ were known. The ref. \cite{KGorska11a} provides an explicit solution for all rational values of admissible parameters, which include \textit{all} integer $M$. We remark that $p_{4}(x, t)$ and $p_{6}(x, t)$ play an important role \cite{KGorska13} in a theory of energy correlations in the ensembles of Hermitian random matrices \cite{EBrezin98a, EBrezin98b} and are related to the problem of phase transitions in chiral QCD models \cite{RAJanik99}.

Let us now consider the two-sided (bilateral) Laplace transform \cite{GDoetsch74, DVWidder46} of $p_{M}(x, t)$, compare Eq. (\ref{E6}). For $t > 0$ and $c \neq 0$ the integral (see Eqs. \eqref{E8} and \eqref{E9})
\begin{equation}\label{E15}
\int_{-\infty}^{\infty} e^{- s y} p_{M}(y, t) dy = e^{\kappa_{M} t s^{M}} 
\end{equation}
is converging absolutely. The validity of Eq. (\ref{E15}) is easy to demonstrate for two cases: $M=2$, where we use formula 2.3.15.11 on p. 344 of \cite{APPrudnikov-1}, and $M=3$, that is proved in \cite{DVWidder79}. For the arbitrary integer $M \geq 2$ the absolute convergence of Eq. (\ref{E15}) is ensured by the appropriate asymptotic behavior of $g_{M}(x)$ at infinity, namely $g^{\rm a}_{M}(x) \approx \exp(- A x^{\beta})$, $\beta \geq 1$, see Eqs. (\ref{e3}) and (\ref{e4}) for $t=1$. 

Eq. (\ref{E15}) is a crucial formula of our paper because by making the substitution $s = \partial_{x}$ we obtain 
\begin{equation}
\hat{U}_{M}(t) = \int_{-\infty}^{\infty} \exp(- y \ulamek{\partial}{\partial x}) p_{M}(y, t) dy,
\end{equation}
where $\exp(- y \ulamek{\partial}{\partial x})$ is the shift operator. That gives the integral representation of the evolution operator of Eq.~(\ref{E3}) and, as a consequence, the general form of $F_{M}(x, t)$: 
\begin{align}\label{E16}
F_{M}(x, t) &= \int_{-\infty}^{\infty} \exp(- y \ulamek{\partial}{\partial x}) p_{M}(y, t) f(x) dy \nonumber\\
&= \int_{-\infty}^{\infty} p_{M}(y, t) f(x-y) dy.
\end{align}
Eq. (\ref{E16}) for $M=2$ is the Gauss-Weierstrass transform, see \cite{DVWidder75}, and for $M=3$ it is the Airy Ai transform, see \cite{DVWidder79, DBabusci11, GDattoli07}. Several examples of $f(x)$ such that Eq. (\ref{E16}) can be evaluated analytically are given in Sec. V.

\section{The signed L\'{e}vy stable laws}\label{Levy}

This Section is devoted to the exact and explicit forms of $g_{2m}(u)$ for $m = 1, 2, 3, \ldots$, see Eq. (\ref{E13}), and thereafter we look closer at their asymptotic behavior at infinity and the associated Hamburger moment problem\footnote{For the purpose of this paper we use this terminology for signed functions.}.

In the spirit of Refs. \cite{KAPenson10, KGorska11a} and \cite{KGorska13} we shall provide the exact expression of $g_{2m}(u)$ using the Mellin transform.

We start with supposing that for certain values of complex $s$ the Mellin transform of $g_{2m}(u)$ exists:
\begin{equation}\label{E17}
g^{\star}_{2m}(s)  = \mathcal{M}[g_{2m}(u), s] = \int_{0}^{\infty} u^{s-1} g_{2m}(u) du,
\end{equation}
and $g_{2m}(u) = \mathcal{M}^{-1}[g^{\star}_{2m}(s), u]$. Then, using Eq. (\ref{E13}) and \cite[see formulas 2.5.3.10 on p. 387 and 2.3.3.1 on p. 322]{APPrudnikov-1} we have
\begin{equation}\label{E18}
g^{\star}_{2m}(s)  = \frac{1}{2\pi m} \Gamma(s) \Gamma\left(\frac{1-s}{2m}\right) \cos\left(\frac{s \pi}{2}\right).
\end{equation}
With the help of the second Euler's reflection formula we express the cosine via gamma function. Inverting the Mellin transform of $g^{\star}_{2m}(s)$ we obtain 
\begin{align}\label{E19}
g_{2m}(u) &= \mathcal{M}^{-1}[g^{\star}_{2m}(s), u] \nonumber\\
&= \frac{1}{2\pi i} \int_{L} u^{-s} \frac{\Gamma(s-1) \Gamma\left(1 - \frac{s-1}{2m}\right)}{ \Gamma\left(\frac{s-1}{2}\right) \Gamma\left(1 - \frac{s-1}{2}\right)} ds,
\end{align}
with the contour $L$ lying between the poles of $\Gamma(s-1)$ and those of $\Gamma\Big(1 - \ulamek{s-1}{2m}\Big)$. After applying the Gauss-Legendre multiplication formula to the gamma functions in Eq. (\ref{E19}) we can express $g_{2m}(u)$ in terms of the Meijer G functions $G^{m, n}_{p, q}\left(x\Big\vert {(\alpha_{p}) \atop (\beta_{q})}\right)$ \cite{APPrudnikov-3} 
\begin{equation}
g_{2m}(u) = \sqrt{\frac{m}{\pi}}\, \frac{1}{u} G^{\,\,\,\,1, \,\,\, 2m}_{3m, m +1}\left(\frac{(2m)^{2m}}{u^{2m}}\Big\vert {\Delta(2m, 0), \Delta(m, 0) \atop 0, \Delta(m, 0)}\right), 
\end{equation}
where $\Delta(k, a) = \ulamek{a}{k}, \ulamek{a+1}{k}, \ldots, \ulamek{a + k - 1}{k}$ is a special list of $k$ elements. Furthermore, it turns out that $g_{2m}(u)$ is a finite sum of $m$ generalized hypergeometric functions of type ${_{p}F_{q}}\left({(\alpha_{p}) \atop (\beta_{q})}\Big| z \right)$:
\begin{equation}\label{E20}
g_{2m}(u) = \sum_{j=1}^{m} \frac{c_{j}(m)}{u^{2-2j}} \,{_{2}}F_{2 m}\left({1, 1 + \ulamek{2j-1}{2m} \atop \Delta(2m, 2j)} \Big\vert z \right)
\end{equation} 
with $z = (-1)^{m} [u/(2m)]^{2m}$ and coefficients $c_{j}(m)$ read
\begin{eqnarray}\label{E21}
c_{j}(m) &=& \frac{\sqrt{m/\pi}}{(2m)^{2j-1} \pi^{m}} \Gamma\left(1 + \frac{2j-1}{2m}\right) \nonumber \\[0.4\baselineskip]
&\times& \frac{\left[\prod_{i=1}^{2j-1} \Gamma\left(\ulamek{i-2j}{2m}\right)\right]\, \left[\prod_{i=2j+1}^{2m} \Gamma\left(\ulamek{i-2j}{2m}\right)\right]}{\left[\prod_{i=0}^{m-1} \sin\left(\pi \ulamek{i}{m} - \pi \ulamek{2j-1}{2m}\right)\right]^{-1}}.
\end{eqnarray}
The formulas Eq. (\ref{E20}) and (\ref{E21}) follow from the Eq. 8.2.2.3 of \cite{APPrudnikov-3} and are in agreement with Eq. (2.4) of \cite{DTHaimo92}. Here, we have used the compact notation of ${_{p}F_{q}}$'s functions where the upper (lower) list of parameters corresponds to the first (second) list of parameters in standard notation. We remark that in the lists of parameters of ${_{2}F_{2m}}$ two cancellations of the same terms appear due to the obvious identity ${_{p+r} F_{q+r}}\left({(\alpha_{p}), (\gamma_{r}) \atop (\beta_{p}), (\gamma_{r})}\right) = {_{p} F_{q}}\left({(\alpha_{p}) \atop (\beta_{p})}\right)$, where $(\gamma_{r})$ is an arbitrary sequence of $r$ parameters. Thus Eq. (\ref{E20}) finally reads as the sum of $m$ generalized hypergeometric functions of type ${_{0}F_{2m-2}}\left({-\atop (\beta_{2m-2})}\Big\vert z\right)$ \cite{DTHaimo92}.

Formulas (\ref{E20}) and (\ref{E21}) reconstruct the explicitly known cases presented in \cite{KGorska11a} and give an unlimited number of new exact solutions $g_{2m}(u)$, \textit{e.g.} for $m = 4$
\begin{eqnarray}\label{E22}
g_{8}(u) & =& c_{4}(4) u^{6}\, {_{0}F_{6}}\left({- \atop \ulamek{9}{8}, \ulamek{5}{4}, \ulamek{11}{8}, \ulamek{3}{2}, \ulamek{13}{8}, \ulamek{7}{4}}\Big\vert \frac{u^{8}}{8^{8}}\right) \nonumber \\[0.4\baselineskip]
&+& c_{3}(4) u^{4}\, {_{0}F_{6}}\left({- \atop \ulamek{3}{4}, \ulamek{7}{8}, \ulamek{9}{8}, \ulamek{5}{4}, \ulamek{11}{8}, \ulamek{3}{2}}\Big\vert \frac{u^{8}}{8^{8}}\right) \nonumber \\[0.4\baselineskip]
&+& c_{2}(4) u^{2}\, {_{0}F_{6}}\left({- \atop \ulamek{1}{2}, \ulamek{5}{8}, \ulamek{3}{4}, \ulamek{7}{8}, \ulamek{9}{8}, \ulamek{5}{4}}\Big\vert \frac{u^{8}}{8^{8}}\right) \nonumber \\[0.4\baselineskip]
&+& c_{1}(4) \, {_{0}F_{6}}\left({- \atop \ulamek{1}{4}, \ulamek{3}{8}, \ulamek{1}{2}, \ulamek{5}{8}, \ulamek{3}{4}, \ulamek{7}{8}}\Big\vert \frac{u^{8}}{8^{8}}\right).
\end{eqnarray}
The coefficients $c_{j}(4)$ for $j=1, \ldots, 4$ in Eq. (\ref{E22}) are equal to $\sqrt{2} \cos\big(\ulamek{\pi}{8}\big)/[4 \Gamma\big(\ulamek{7}{8}\big)]$, $-\sqrt{2} \sin\big(\ulamek{\pi}{8}\big)/[8 \Gamma\big(\ulamek{5}{8}\big)]$, $\sqrt{2}\sin\big(\ulamek{\pi}{8}\big) \sin\big(\ulamek{3\pi}{8}\big) \Gamma\big(\ulamek{5}{8}\big)/(96\pi)$, and $-\sqrt{2}\sin\big(\ulamek{3\pi}{8}\big) \sin\big(\ulamek{\pi}{8}\big) \Gamma\big(\ulamek{7}{8}\big)/(2880 \pi)$, respectively.
\begin{figure}[!h]
\begin{center}
\includegraphics[scale=0.44]{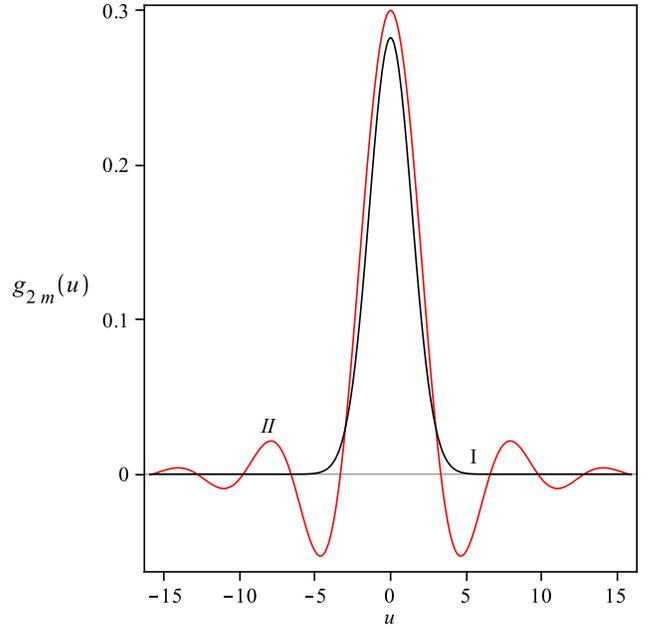}
\caption{\label{fig1} (Color online) Plot of the function $g_{2m}(u)$ for $m=1$ and $m=4$; \textit{i.e.} $g_{2}(u) = \exp(-u^{2}/4)/(2\sqrt{\pi})$ (I; black line) and $g_{8}(u)$ (II; red line) given in Eq. (\ref{E22}).}
\end{center}
\end{figure}

In Fig.~\ref{fig1} $g_{8}(u)$ (I; red line) and $g_{2}(u)$ (II; black line) are presented. The tails of symmetric function $g_{8}(u)$ oscillate. The amplitude of these oscillations is decreasing with increasing values of $|u|$. The analogous behavior is observed in the other examples of $g_{2m}(u)$, for instance see Fig. 1 in \cite{KGorska13} where functions $g_{4}(u)$ and $g_{6}(u)$ are exhibited. Figs.~1 and 2 in \cite{KGorska13} well illustrate the considerations presented in \cite{XLi93}. 

\subsection{Asymptotics of $g_{2m}(u)$.}

The asymptotics of $g_{4}(u)$ and $g_{6}(u)$ for large $u$ are presented in \cite{KGorska13, TMGaroni02}, whereas the general formula of $g^{\rm a}_{2m}(u)$ ($m=1, 2, \ldots$) for large $u$ is given in \cite{TGaroni03, ALachal03}. Here, following \cite{TGaroni03, ALachal03}, we just present their compact form:
\begin{eqnarray}\nonumber
g^{\rm a}_{2m}(u) &\sim& \frac{\sqrt{2} (2m)^{-\ulamek{1}{4m-2}}}{\sqrt{(2m-1)\pi}} u^{-\ulamek{m-1}{2m-1}} e^{-\sin\Big[\ulamek{\pi}{2(2m-1)}\Big] Z} \\[0.4\baselineskip] 
&\times& \cos\left\{\cos\Big[\ulamek{\pi}{2(2m-1)}\Big] Z - \frac{m-1}{2m-1} \frac{\pi}{2}\right\}, \nonumber
\end{eqnarray} 
$Z = (2m-1)[u/(2m)]^{\frac{2m}{2m-1}}$, which guarantees the absolute convergence of (\ref{E15}).
For small $u$, the L\'{e}vy signed function can be represented by the series
\begin{equation}
\tilde{g}^{\rm a}_{2m}(u) \sim \frac{1}{2m \pi} \sum_{r=0}^{\infty} \frac{(-1)^{r}}{(2r)!} \Gamma\left(\frac{r}{m} + \frac{1}{2m}\right) u^{2r},
\end{equation}
which has the infinite radius of convergence.

\subsection{Hamburger moment problem for signed $g_{2m}(u)$.}

Let us consider the following integral defining Hamburger moments of $g_{M}(u)$
\begin{equation}\label{E23}
h_{M}(\mu) = \int_{-\infty}^{\infty} u^{\mu} g_{M}(u) du, 
\end{equation}
for $M = 2, 3, \ldots$ and arbitrary real $\mu$. In this Section we will look closer at the case of even $M = 2m$, whereas Eq. (\ref{E23}) for odd $M = 2m +1$ will be studied in Sec. IV below. 

Since $g_{M}(u)$ is an even function for $M=2m$, the integral (\ref{E23}) vanishes by symmetry for all odd $\mu = (2n +1)$, regardless of the value of $m$ ($m=1, 2, \ldots$). The calculation of $h_{2m}(2n)$ is a somewhat subtler problem which has been carefully analyzed in \cite{TMGaroni02}. Decomposing $h_{2m}(\mu)$ into two symmetric parts, employing Eq. (\ref{E18}) and making some simple transformation, we express Eq. (\ref{E23}) for $M=2m$, ($m=1, 2, \ldots$), in the form, compare Eq. (\ref{E17}):
\begin{align}
h_{2m}(\mu) &= [1 + (-1)^{\mu}] g^{\star}_{2m}(\mu+1) \nonumber\\
&= \frac{1}{2 m} [1 + (-1)^{\mu}] \frac{\Gamma\big(1+\mu\big)}{\Gamma\big(1 + \ulamek{\mu}{2m}\big)} \frac{\sin\big(\ulamek{\pi\mu}{2}\big)}{\sin\big(\ulamek{\pi\mu}{2m}\big)}.
\end{align}
The moments $h_{2m}(\mu)$ vanish for $\mu = 2 (m p + r)$, $p~=~1, 2, \ldots$, $r = 1, 2, \ldots, m-1$. The only non-zero terms of $h_{2m}(\mu)$ occur for $\mu = 2mp$ for which
\begin{equation}\label{E24}
\lim_{\mu\to 2mp}h_{2m}(\mu) = (-1)^{(1+m)p} \frac{(2 m p)!}{p!}.
\end{equation}
We see that $h_{2m}(0) = 1$, that is $g_{2m}(u)$ is normalized to unity. The first few non-zero terms of $h_{2m}(\mu)$ for $m~=~1, 2, 3$ and $\mu = 0, 1, \ldots, 12$ are presented in Tab. \ref{tab1}. 
\setlength{\tabcolsep}{.4mm}
   \renewcommand\arraystretch{1.5}
\begin{table}[!h]
\begin{center}
\begin{tabular}{c | c c c c c c c c c c c c c}
$\mu$ & $\,\,0$ & $\,\,1$ & $\,\,2$ & $\,\,3$ & 4 & $\,\,5$ & 6 & $\,\,7$ & 8 & $\,\,9$ & 10 & 11 & 12 \\ \hline
$h_{2}(\mu)$ & 1 &  & 2 &  & 12 &  & 120 &  & 1680 &  & 30240 &  & 665280 \\
$h_{4}(\mu)$ & 1 &  &  &  & -24 &  &  &  & 20160 &  &  &  & -79833600 \\
$h_{6}(\mu)$ & 1 &  &  &  &  &  & 720 &  &  &  &  &  & 239500800 
\end{tabular}
\caption{\label{tab1} The values of the integrals (\ref{E23}) for $M = 2, 4, 6$ and $\mu = 0, 1, \ldots, 12$.}
\end{center}
\end{table}

\section{Generalized Airy Functions}\label{Airy}

To derive the exact and explicit form of $Ai^{(2m+1)}(u)$ we apply the method od Sec. \ref{Levy} with the Mellin transform $[Ai^{(2m+1)}(s)]^{\star} = \mathcal{M}[Ai^{(2m+1)}(u); s]$ for complex $s$. Then, using Eq. (\ref{E14}) and formulas 2.5.3.10 on page 387 and 2.3.3.1 on page 322 of \cite{APPrudnikov-1}, we get
\begin{equation}
\Big[Ai^{(2m+1)}(s)\Big]^{\star} = \frac{\Gamma(s) \Gamma(\ulamek{1-s}{2m+1})}{\pi(2m+1)} \cos\left[\frac{\pi (2ms +1)}{2(2m+1)}\right]
\end{equation}
and performing the Mellin inversion as in Eq. (\ref{E19}), we obtain the exact and explicit expression for $Ai^{(2m+1)}(u)$ in terms of Meijer G function:
\begin{align}
&Ai^{(2m+1)}(u) = \sqrt{\frac{2m+1}{2\pi}}\, \frac{1}{u} \nonumber\\
&\quad\times G^{\,\,\,\,1, \,\,\, 2m+1}_{3m+1, m +1}\left(\frac{(-1)^{m}}{\tilde{z}}\Big\vert {\Delta(2m+1, 0), \Delta(m, 0) \atop 0, \Delta(m, 0)}\right)
\end{align}
with $\tilde{z} = (-1)^{m}[u/(2m+1)]^{2m+1}$. Furthermore, the Meijer G function is converted to the finite sum of the generalized hypergeometric functions: 
\begin{equation}\label{E25}
Ai^{(2m + 1)}(u) = \sum_{j=1}^{2m} \frac{b_{j}(m)}{u^{1-j}} {_{2}F}_{2m + 1}\left({1, 1 + \ulamek{j}{2m +1} \atop \Delta(2m+1, 1+j)} \Big\vert \tilde{z} \right)
\end{equation}
with 
\begin{eqnarray}\label{E26}
b_{j}(m) &=& \frac{\sqrt{(2m+1)}}{(2m+1)^{j} \pi^{m+1}}\, \Gamma\left(1 + \frac{j}{2m+1}\right) \nonumber\\[0.4\baselineskip]
&\times& \frac{\left[\prod_{i=1}^{j} \Gamma\big(\ulamek{i-j-1}{2m+1}\big)\right]\, \left[\prod_{i=j+2}^{2m+1} \Gamma\big(\ulamek{i-j-1}{2m+1}\big)\right]}{\left[\prod_{i=0}^{m} \sin\big(\pi \ulamek{i}{m+1} - \pi\ulamek{j}{2m+1}\big)\right]^{-1}}. 
\end{eqnarray}
In obtaining Eqs. (\ref{E25}) and (\ref{E26}) we have again used the formula 8.2.2.3 of \cite{APPrudnikov-3}. We point out that similarly as in the case of the L\'{e}vy signed functions of the previous section, in the list of parameter ${_{2}F}_{2m+1}$ the cancellation of the same two terms occurs. That gives the finite sum of $2m$ hypergeometric functions of type ${_{0}F}_{2m-1}\left({- \atop (\beta_{2m-1})}\Big\vert \tilde{z}\right)$. Eqs. (\ref{E25}) and (\ref{E26}) are in agreement with the case of $t=1$ of the formula for $u_{2n + 1}(x, t)$ on page 2 of \cite{EOrsinger12}. 

The formulas (\ref{E25}) and (\ref{E26}) reproduce the well-known case of conventional Airy Ai function (see Eqs. (\ref{E25}) and (\ref{E26}) for $m=1$) and give the exact and explicit form of the generalized Airy Ai functions $Ai^{(2m+1)}(u)$ which were only numerically obtained in \cite{RCYChin78}. Without loss of the generality below we are writing out $Ai^{(2m+1)}(u)$ for $m=2$:
\begin{eqnarray}\label{E27}
Ai^{(5)}(u) &=& b_{1}(2)\, {_{0}F}_{3}\left({- \atop \ulamek{2}{5}, \ulamek{3}{5}, \ulamek{4}{5}} \Big\vert \frac{u^{5}}{5^{5}}\right) \nonumber\\
&+& b_{2}(2)\, u\, {_{0}F}_{3}\left({- \atop \ulamek{3}{5}, \ulamek{4}{5}, \ulamek{6}{5}} \Big\vert \frac{u^{5}}{5^{5}}\right) \nonumber \\
& +& b_{3}(2)\, u^{2} {_{0}F}_{3}\left({- \atop \ulamek{4}{5}, \ulamek{6}{5}, \ulamek{7}{5}} \Big\vert \frac{u^{5}}{5^{5}}\right) \nonumber\\
&+& b_{4}(2)\, u^{3} {_{0}F}_{3}\left({- \atop \ulamek{6}{5}, \ulamek{7}{5}, \ulamek{8}{5}} \Big\vert \frac{u^{5}}{5^{5}}\right).
\end{eqnarray}
The coefficients $b_{j}(2)$, $j=1, \ldots, 4$, are equal to $\sqrt{5}A /[10 \Gamma(\ulamek{4}{5})\sin(\ulamek{\pi}{5})]$, $-\sqrt{5}B/[10 \Gamma(\ulamek{3}{5}) \sin(\ulamek{2\pi}{5})]$, $-\sqrt{5} B \Gamma(\ulamek{3}{5})/(20 \pi)$, and $\sqrt{5} A \Gamma(\ulamek{4}{5}) /(60\pi)$, where $A = \sin(\ulamek{3\pi}{10})/\sin(\ulamek{2\pi}{5})$ and $B= \sin(\ulamek{\pi}{10})/\sin(\ulamek{\pi}{5})$, respectively. 

\begin{figure}[!h]
\begin{center}
\includegraphics[scale=0.44]{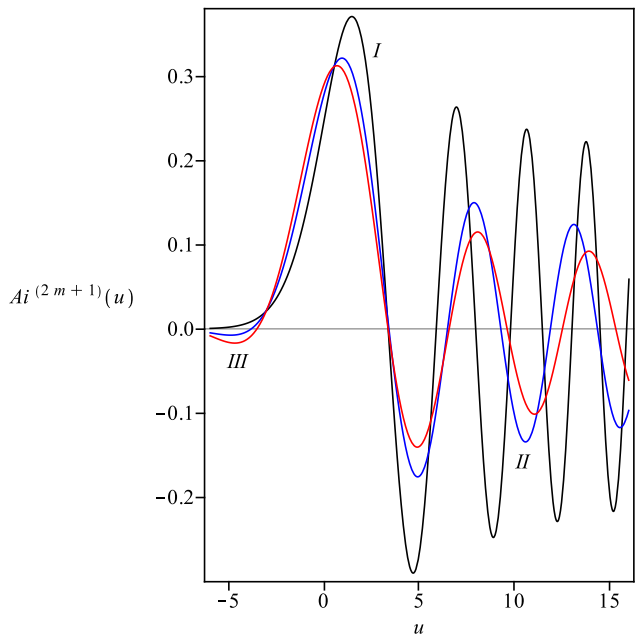}
\caption{\label{fig2} (Color online) Plot of the functions $Ai^{(2m+1)}(u)$ for $m = 1, 2, 3$; \textit{i.e.} $Ai^{(3)}(u) = 3^{-1/3}Ai(-u/3^{-1/3})$ (I; black line) and $Ai^{(5)}(u)$ (II; blue line) given in Eq. (\ref{E27}), and $Ai^{(7)}(u)$ (III; red line) calculated from Eqs. (\ref{E25}) and (\ref{E26}) for $m=3$, respectively. Note that $Ai^{(3)}(x)$ is not oscillating for $x < 0$.}
\end{center}
\end{figure}

In Fig. \ref{fig2} the functions $Ai^{(2m+1)}(u)$ are presented from Eqs. (\ref{E25}) and (\ref{E26}) for $m=1$ (I; black line), $m=2$ (II; blue line), and $m=3$ (III; red line). For these generalized Airy Ai functions we observe the oscillation for positive \textit{and} negative $u$, which constitutes a good illustration of considerations presented in \cite{ALachal03}.  

\begin{figure}[!h]
\begin{center}
\includegraphics[scale=0.39]{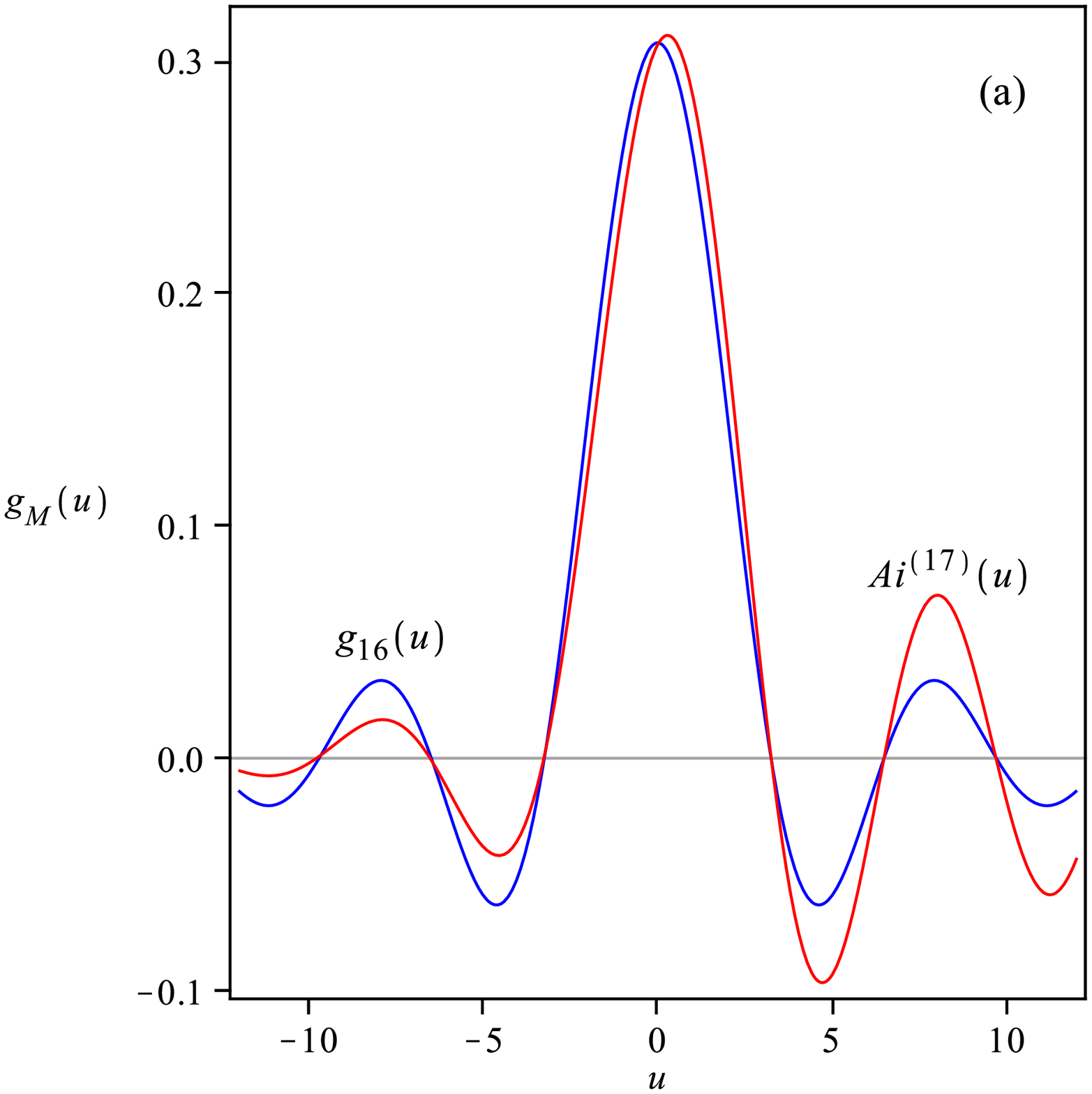}
\includegraphics[scale=0.38]{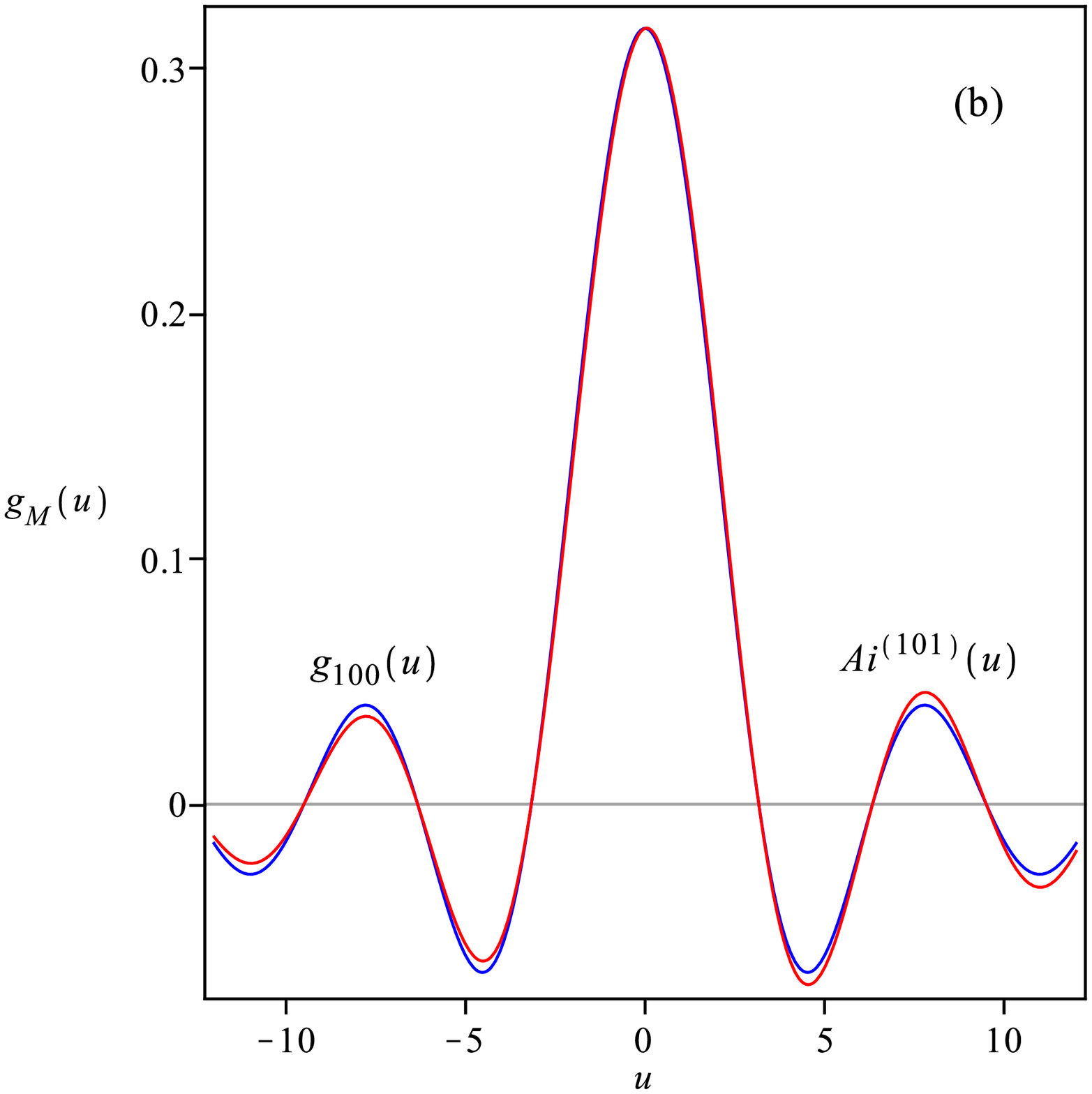}
\caption{\label{fig2a} (Color online) Comparison of the functions $g_{2m}(u)$ (blue line) and $Ai^{(2m+1)}(u)$ (red line) for given $m$. In Fig. \ref{fig2a}a we present the case $m=8$, whereas in Fig. \ref{fig2a}b we present the case $m=50$. It is seen that for large $m$ the difference between even and odd case becomes negligible.}
\end{center}
\end{figure}

The comparison between functions $g_{2m}(u)$ and $Ai^{(2m+1)}(u)$ for fixed values of $m$ is shown in Figs. \ref{fig2a}; on Fig. \ref{fig2a}a it is done for $m=8$, whereas the Fig. \ref{fig2a}b is for $m=50$. It turns out that for large $m$ the difference between $g_{2m}(u)$ and $Ai^{(2m+1)}$ are negligible.

\subsection{Asymptotics of $Ai^{(2m + 1)}(u)$.}

The asymptotic expansions of the generalized Airy Ai functions for large negative and positive values of argument are given as
\begin{eqnarray}\label{E28}
Ai^{(2m+1)}_{\rm a^{-}}(u) &\sim& \frac{1}{\sqrt{m\pi}} \left[(2m+1) |u|^{2m-1}\right]^{-\ulamek{1}{4m}} e^{- \sin\big(\ulamek{\pi}{2m}\big) |\tilde{Z}|} \nonumber\\[0.4\baselineskip]
&\times&\cos\left[\cos\big(\ulamek{\pi}{2m}\big) |\tilde{Z}| - \frac{m-1}{m} \frac{\pi}{4}\right],
\end{eqnarray}
for $u\to-\infty$, and 
\begin{equation}\label{E29}
Ai^{(2m+1)}_{\rm a^{+}}(u) \sim \frac{1}{\sqrt{m\pi}} \left[(2m+1) u^{2m-1}\right]^{-\ulamek{1}{4m}} \cos\left(\tilde{Z} - \frac{\pi}{4}\right),
\end{equation}
for $u\to\infty$; $\tilde{Z} = 2m [u/(2m+1)]^{\frac{2m+1}{2m}}$, see Proposition 2 in \cite{ALachal03}. The symbol $Ai^{(2m+1)}_{\rm a^{-}}(u)$ [$Ai^{(2m+1)}_{\rm a^{+}}(u)$] denotes the asymptotic estimate for large negative (positive) $u$. For $m=1$ Eqs. (\ref{E28}) and (\ref{E29}) lead to the known formulas on the asymptotic behavior of $Ai^{(3)}(u)$, see \textit{e.g.} \cite{DVWidder79}. For $m=2$ we have
\begin{eqnarray}\nonumber
Ai^{(5)}_{\rm a^{-}}(u) &\sim& \frac{5^{-1/8}}{\sqrt{2 \pi}} |u|^{-3/8} e^{- |\tilde{Z}|} \sin\left(\frac{\sqrt{2}}{2} |\tilde{Z}| + \frac{3\pi}{8}\right), \nonumber \\[0.4\baselineskip]
Ai^{(5)}_{\rm a^{+}}(u)&\sim& \frac{5^{-1/8}}{\sqrt{2 \pi}} u^{-3/8} \sin\left(\tilde{Z} + \frac{\pi}{4}\right), 
\end{eqnarray}
with $\tilde{Z} = 4 \big(u/5\big)^{5/4}$. We point out that for large values of $m$, $Ai^{(2m+1)}_{\rm a^{-}}(u)$ approaches $Ai^{(2m+1)}_{\rm a^{+}}(u)$.

The asymptotic behavior for small values of $u$ is obtained by using the Taylor expansion of $Ai^{(2m+1)}(u)$ or formula 2.5.21.6 on p. 430 of \cite{APPrudnikov-1}. That gives
\begin{align}\label{E30}
\tilde{Ai}^{(2m+1)}_{\rm a}(u) &\sim \frac{\pi^{-1}}{(2m+1)}  \sum_{r=0}^{\infty} \frac{u^{r}}{r!} \Gamma\left(\frac{1+r}{2m+1}\right) \nonumber\\
&\times \cos\left(\frac{1-2m r}{4m+2} \pi\right).
\end{align}
The series in Eq. (\ref{E30}) has the infinite radius of convergence. For a given $m$ the summation can be carried out and it agrees with the general formula of Eq. (\ref{E25}).

\subsection{Hamburger moment problem for $Ai^{(2m + 1)}(u)$.}

Now, we look closer at the Hamburger moment problem of $Ai^{(2m+1)}(u)$. For $M=2m+1$, Eq. (\ref{E23}) decomposes into two integrals according to the sign of $u$. That gives the sum
\begin{equation}
h_{2m+1}(\mu) = g^{+, \star}_{2m+1}(\mu+1) +  (-1)^{\mu}\, g^{-, \star}_{2m+1}(\mu+1),
\end{equation}
where $g^{\pm, \star}_{2m+1}(s)$ denotes the Mellin transform of $g^{\pm}_{2m+1}(u)$ given in Eq. (\ref{E14}). Considering separately the case of even and odd moments, after using Eqs. (\ref{E17}), (\ref{E14}) and formula 2.3.3.1 on page 322 of \cite{APPrudnikov-1} we have
\begin{equation}\label{E31}
h_{2m+1}(2n) = \frac{(2n-1)!}{\Gamma\Big(\ulamek{2n}{2m+1}\Big)} \frac{\sin(\pi n)}{\sin\Big(\ulamek{\pi n}{2m+1}\Big)}, 
\end{equation}
\begin{equation}\label{E32}
h_{2m+1}(2n+1) = -\frac{(2n)!}{\Gamma\Big(\ulamek{2n+1}{2m+1}\Big)} \frac{\sin(\pi n)}{\cos\Big(\ulamek{\pi}{2} \ulamek{2n+1}{2m+1}\Big)}.
\end{equation}

Let us analyze properties of Eqs. (\ref{E31}) and (\ref{E32}). At first we see that $h_{2m+1}(2n)$ vanish for integer $n$ except for $n$ being the multiple of $(2m+1)$. For $n = (2m+1)k$, $k = 0, 1, \ldots$ the ratio of sines in $h_{2m+1}(2n)$ is finite and it goes to $(2m+1)$. The only non-zero terms can be written in the form
\begin{equation}\label{E33}
\lim_{n\to (2m+1) k} h_{2m+1}(2n) = \frac{[(2m+1)2k]!}{(2k)!}.
\end{equation}
It is obvious that $h_{2m+1}(0) = 1$ for $k=0$. Analogical situation as in the above case appears for $h_{2m+1}(2n+1)$ for which for $n = (2m+1)k + m$ the ratio of sine and cosine goes to $(-1)^{m+1} (2m+1)$. That gives
\begin{equation}\label{E34}
\lim_{n\to (2m+1)k + m} h_{2m+1}(2n+1) = (-1)^{m} \frac{[(2m+1)(2k+1)]!}{(2k+1)!}.
\end{equation}

The first few values of non-vanishing terms of $h_{2m+1}(\mu)$ for $m= 1, 2$ and $3$ are presented in Tab. \ref{tab2}. \\
\setlength{\tabcolsep}{.4mm}
   \renewcommand\arraystretch{1.5}
\begin{table}[!h]
\begin{center}
\begin{tabular}{c | c c c c c c c c c c c }
$\mu$ & $\,\,0$ & $\,\,1$ & $\,\,2$ & $\,\,3$ & $\,\,4$ & $\,\,5$ & 6 & $\,\,7$ & $\,\,8$ & 9 & 10  \\ \hline
$h_{3}(\mu)$ & 1 &  &  & -6 &  &  & 360 &  &  & -60480 &   \\
$h_{5}(\mu)$ & 1 &  &  &  &  & 120 &  &  &  &  & 1814400  \\
$h_{7}(\mu)$ & 1 &  &  &  &  &  &  & -5040 &  &  &  
\end{tabular}
\caption{\label{tab2} The values of the integrals (\ref{E23}) for $M = 3, 5, 7$ and $\mu = 0, 1, \ldots, 10$.}
\end{center}
\end{table}

\section{Specific examples}\label{SExp}

The content of this Section concerns the formal deliberations on the relations between $F_{M}(x, t)$, for fixed initial condition, expressed by Eqs. (\ref{E5})  and (\ref{E16}), and the integral transform approach.
\ \\

\noindent
\textbf{(A)} First we observe that Eq. (\ref{E16}) with $f(x) = x^{n}$ and the kernel given in Eq. (\ref{E11}) lead to
\begin{equation}\label{E35}
F^{(n)}_{M}(x, t) = \sum_{k=0}^{n} \left({n}\atop{k}\right) (-1)^{k} x^{n-k} t^{k/M} h_{M}(k),
\end{equation}
where $h_{M}(k)$ are defined in (\ref{E23}) and for the initial condition we have used the Newton's binomial theorem. Without loss of generality, let us  look at the first few terms of (\ref{E35}) for $M=3$. Using to that purpose $h_{3}(k)$ exhibited in Tab. \ref{tab2} we get: $F_{3}^{(0)}(x, t) = 1$, $F_{3}^{(1)}(x, t) = x$, $F_{3}^{(2)}(x, t) = x^{2}$, and $F_{3}^{(3)}(x, t) = x^{3} + 6 t$. That reproduces the Hermite-Kamp\'{e} de F\'{e}ri\'{e}t polynomials $H^{(3)}_{n}(x, t)$ for $n=0, \ldots, 3$. In the general case, from Tabs. \ref{tab1} and \ref{tab2} it emerges that only $[n/M]$th terms of $h_{M}(\mu)$ are different from zero. After changing the summation index in Eq. (\ref{E35}) as $k = Mp$ ($p=0, 1, \ldots, [n/M]$) and employing formulas (\ref{E24}), (\ref{E33}) and (\ref{E34}) we convert Eq. (\ref{E35}) into Eq. (\ref{E4}). Making similar consideration for the initial condition formally written in the form of the power series we can express the representation of $F_{M}(x, t)$ in the form of Eq. (\ref{E5}).

Our approach suggests a new way of looking at the Hermite-Kamp\'{e} de F\'{e}ri\'{e}t polynomials, which according to Eq. (\ref{E35}) can be defined by the operational form
\begin{equation}
H^{(M)}_{n}(x, \kappa_{M} t) = (x - t^{1/M} \hat{h}_{M})^{n} \varphi_{0},
\end{equation}
with
\begin{equation}
\hat{h}_{M}^{k} \varphi_{0} = h_{M}(k),
\end{equation}
where $\varphi_{0}$ is the 'vacuum' and $h_{M}(k)$ are given by Eqs. (\ref{E24}), (\ref{E33}) and (\ref{E34}). See also \cite{GDattoli13} for related considerations in the context of lacunary Laguerre polynomials.

\ \\
\noindent
\textbf{(B)} The next example, which shows the validity of the method based on the integral transform (\ref{E16}), we consider the generalization of the Glaisher formula whose original form reads:
\begin{equation}\label{e1}
F_{2}(x, t) = \exp\left(\kappa_{2} t \frac{\partial^{2}}{\partial x^{2}}\right) e^{-\alpha x^{2}} = \frac{\exp(-\frac{\alpha x^{2}}{1 + 4\alpha^{2} t})}{\sqrt{1+4\alpha^{2} t}}.
\end{equation}
for $t > -1/(4\alpha^{2})$ and $\kappa_{2} = 1$. For that purpose we take as the initial condition $f(x) = g_{M}(\alpha x)$, $\alpha > 0$. For such a choice, we get
\begin{align}\label{E38}
F_{M}(x, t) &= \exp\left(\kappa_{M} t \frac{\partial^{M}}{\partial x^{M}}\right) g_{M}(\alpha x) \nonumber\\
& = \left(1 + \alpha^{M} t\right)^{-1/M} g_{M}\left[\frac{\alpha x}{(1 + \alpha^{M} t)^{1/M}}\right],
\end{align}
$-\infty < x < \infty$ and $t > - \alpha^{-M}$. The constant $\kappa_{M}$ is hidden in the definition of $g_{M}(u)$ given in Eq. (\ref{E12}). Eq. (\ref{E38}) neatly illustrates the \textit{scaling} character of the time evolution from this initial condition. In certain sense it constitutes a rather far-reaching extension of Glaisher-type relations \cite{GDattoli08}, which for $M=3$ can be compactly written as 
\begin{equation}\label{e2}
F_{3}(x, t) = \exp\left(\kappa_{3} t \frac{\partial^{3}}{\partial x^{3}}\right) Ai(-\alpha x) = \frac{Ai\left[\frac{-\alpha x}{(1 + 3 \alpha^{3} t)^{1/3}}\right]}{(1 + 3 \alpha^{3} t)^{1/3}}
\end{equation}
for $t  > -(3\alpha^{3})^{-1}$ and $\kappa_{3} =1$. The analogous expression for $\kappa_{3} = -1$ can be obtained from Eq. (\ref{e2}) by transforming $\alpha$ to $-\alpha$ with the restriction on $t$: $t < (3 \alpha^{3})^{-1}$. These relations are evocative of the reproducing property of L\'{e}vy laws under the L\'{e}vy transform \cite{KGorska12}. We remind the reader that $Ai(z)$ is not a L\'{e}vy signed function. The conditions on validity of Eqs. (\ref{e1}), (\ref{E38}), and (\ref{e2}) are always satisfied for $t>0$, compare with Eq. (\ref{E10}). Below, we will sketch the proof of Eq. (\ref{E38}) in two independent ways: first by using the integral transform of Eq. (\ref{E16}) and then by the study of the action of the evolution operator on the initial condition $g_{M}(\alpha x)$. At first, we use Eq. (\ref{E16}): 
\begin{equation}\label{E39}
F_{M}(x, t) = \frac{1}{t^{1/M}} \int_{-\infty}^{\infty} g_{M}\left(\frac{y}{t^{1/M}}\right) g_{M}[\alpha(x-y)] dy.
\end{equation}
Substituting Eq. (\ref{E12}) into Eq. (\ref{E39}) we have three integrals to calculate, which after changing the order of integration, can be written as
\begin{eqnarray}\label{E40}
F_{M}(x, t) &=& \Re\left\{\int_{0}^{\infty} e^{\kappa_{M} (i y_{1})^{M}} \frac{dy_{1}}{\pi}\int_{0}^{\infty} e^{\kappa_{M} (i y_{2})^{M} + i \alpha x y_{2}} \frac{dy_{2}}{\pi} \right. \nonumber \\[0.4\baselineskip]
&\times& \left.\int_{-\infty}^{\infty} e^{i\big(\ulamek{y_{1}}{t^{1/M}} - \alpha y_{2}\big) y} \frac{dy}{t^{1/M}}  \right\}.
\end{eqnarray}
The integral over $y$ in Eq. (\ref{E40}) is equal to $2\pi\delta(y_{1}~-~\alpha t^{1/M} y_{2}\big)$. That simplifies the integration over $y_{1}$ in Eq. (\ref{E40}) and, in consequence, gives formula (\ref{E38}). Otherwise, $F_{M}(x, t)$ can be obtained by employing Eqs. (\ref{E3}) and (\ref{E12}), illustrated below:
\begin{eqnarray}\nonumber
F_{M}(x, t) &=& \Re \left\{\frac{1}{\pi} \int_{0}^{\infty} \exp\left(\kappa_{M} t \frac{\partial^{M}}{\partial x^{M}}\right) e^{i \alpha x y} e^{\kappa_{M}(iy)^{M}} dy\right\} \\[0.4\baselineskip]
&=& \Re \left[\frac{1}{\pi} \int_{0}^{\infty} e^{i \alpha x y + \kappa_{M} (i y)^{M}(1 + \alpha^{M} t)} dy\right]. \nonumber
\end{eqnarray}
After introducing $y = u (1 + \alpha^{M} t)^{-1/M}$ we will recover Eq. (\ref{E38}).

\ \\
\noindent
\textbf{(C)} In the last two examples, we choose, rather arbitrarily, the initial condition given by the Cauchy distribution $f(x) = \ulamek{1}{\pi(\alpha^{2}+x^{2})}$, $\alpha > 0$. For that choice of $f(x)$ and for even $M$ we are able to find the exact and explicit form of $F_{M}(x, t)$. Using Eq. (\ref{E16}) with the integral kernel $p_{2m}(y, t)$ defined in Eqs. (\ref{E11}) and (\ref{E12}), we obtain
\begin{equation}\nonumber
F_{2m}(x, t) = \Re\left\{\int_{-\infty}^{\infty} \frac{du}{\pi^{2}(\alpha^{2} + u^{2})} \left[\int_{0}^{\infty} e^{\frac{i(x-u)z}{t^{1/(2m)}}  - z^{2m}} \frac{dz}{t^{\frac{1}{2m}}}\right] \right\}, 
\end{equation}
where $u = (x-y)$. Calculating at first the integral over $u$ and thereafter using formula 2.3.2.13 of \cite{APPrudnikov-1}, we get
\begin{eqnarray}\label{E41}
&&F_{2m}(x, t)  = \Re\left\{\int_{0}^{\infty} \exp\left[-\frac{\alpha-ix}{t^{1/(2m)}} z - z^{2m}\right] \frac{dz}{\pi \alpha t^{1/(2m)}}\right\} \nonumber \\
&&\quad = \frac{1}{2m\alpha\pi} \Re\left\{\sum_{j=1}^{2m} \frac{(-1)^{j-1}}{(j-1)!} \Gamma\left(\frac{j}{2m}\right) \frac{(\alpha-ix)^{j-1}}{t^{j/(2m)}}\right. \nonumber \\
&& \quad\times \left.{_{2}F_{2m}}\left({1, \ulamek{j}{2m} \atop \Delta(2m, j)}\Big\vert \frac{(\alpha-ix)^{2m}}{t (2m)^{2m}}\right)\right\}.
\end{eqnarray}
For odd $M$ the function $F_{2m+1}(x, t)$ given by
\begin{align}\label{E42}
F_{2m+1}(x, t) &= \frac{1}{\pi \alpha t^{1/(2m+1)}} \int_{0}^{\infty} \exp\left[-\frac{z \alpha}{t^{1/(2m+1)}}\right] \nonumber\\
&\times \cos\left(\frac{xz}{t^{1/(2m+1)}} - z^{2m+1}\right) dz
\end{align}
can be calculated only numerically.

\begin{figure}[!h]
\begin{center}
\includegraphics[scale=0.44]{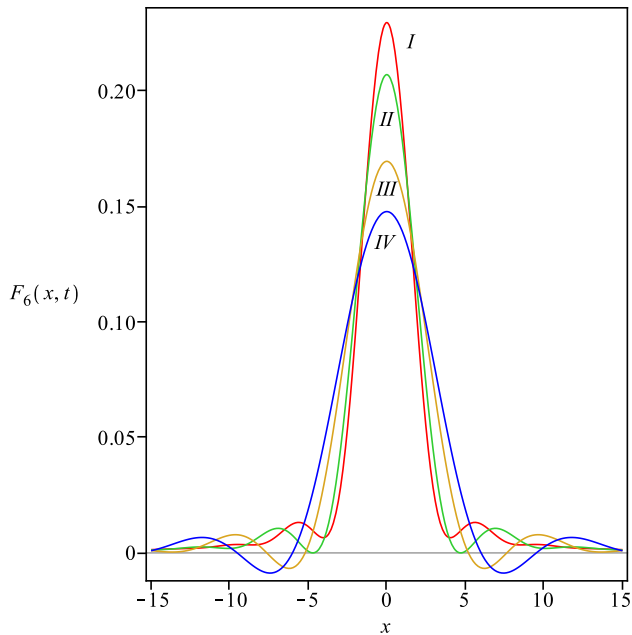}
\caption{\label{fig5} (Color online) Plot of $F_{6}(x, t)$ given by Eq. (\ref{E41}) for $\alpha = 1$ and fixed value of $t$; line I is for $t = 0.125$ (red line), line II is for $t=0.415$ (green line), line III is for $t=3$ (yellow line), and line IV (blue line) is for $t=10$.}
\end{center}
\end{figure}

For $m = 3$ the functions $F_{6}(x, t)$ given in Eq. (\ref{E41}) for $\alpha = 1$ and $t = 0.125, 0.415, 3$ and $10$ are illustrated in Fig.~\ref{fig5}. The calculations indicate that for $\alpha =1$ there exists the border time $t_{1} \cong 0.415$ for which $F_{6}(x, t)$, $t~ <~ t_{1}$, is positive, see line I in Fig. \ref{fig5}. For $t > t_{1}$ the function $F_{6}(x, t)$ is negative, see lines III and IV in Fig. \ref{fig5}. The line II in Fig. \ref{fig5} presents the border case between the two previous situations. The function $F_{6}(x, t_{1})$ is positive with the roots at the points $x_{1} \cong \pm 4.839$. The roots of $F_{6}(x, t\geq t_{1})$ and the depth of negative parts of $F_{4}(x, t>t_{1})$ depend on values of parameter $\alpha$ and they can be minimized for the appropriate choice of $\alpha$ for given~$t$.

\begin{figure}[!h]
\begin{center}
\includegraphics[scale=0.54]{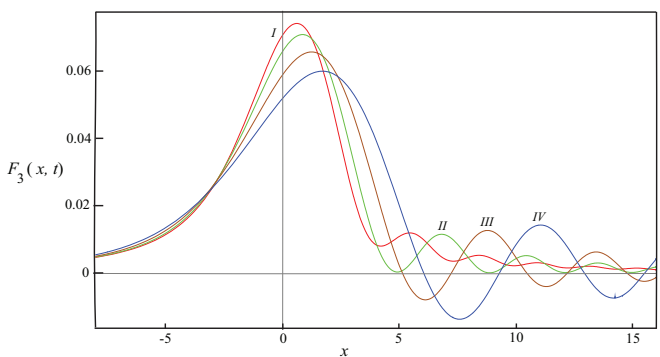}
\caption{\label{fig6} (Color online) Plot of $F_{3}(x, t)$ given in Eq. (\ref{E42}) for $\alpha = 2$ and fixed value of $t$; line I is for $t = 0.5$ (red line), line II is for $t=0.926$ (green line), line III is for $t=2$ (brown line), and line IV (blue line) is for $t=4$.}
\end{center}
\end{figure}

In Fig. \ref{fig6} for $m=1$ the function $F_{3}(x, t)$ is presented, as numerically calculated from Eq. (\ref{E42}) for $\alpha = 2$ and $t=0.5, 0.926, 2$ and $4$. For a given value of $\alpha$, we can also find  the border time $\tilde{t}_{1}$ for which $F_{3}(x, \tilde{t}_{1})$ ceases to be strictly positive and starts to have the roots. For $\alpha = 2$ the border time $\tilde{t}_{1}$ is equal to $0.926$. The strictly positive function $F_{3}(x, t < \tilde{t}_{1})$ is shown as the line I in Fig. \ref{fig6}, $F_{3}(x, \tilde{t}_{1})$ is presented in line II, whereas two signed functions $F_{3}(x, t > \tilde{t}_{1})$ are illustrated in lines III and IV. The existence of the negative parts in Fig. \ref{fig6} are relicts of the oscillations of the integral kernel $p_{3}(x, t)$ and they can be minimized by the suitable choice of $\alpha$ for fixed $t$.

\ \\
\noindent
\textbf{(D)} The formalism developed so far shows a wide flexibility. It can be applied for solving a large class of partial differential equations in the form
\begin{equation}\label{E43}
\frac{\partial}{\partial t} \tilde{F}_{M}(x, t) = \mathcal{\hat{O}}_{M} \tilde{F}_{M}(x, t)
\end{equation}
with $\hat{O}_{M}$ the differential operator of order $M$ being the function of $x$ and $\ulamek{\partial}{\partial x}$ and with the initial condition $\tilde{F}(x, 0) = \tilde{f}(x)$. According to the technique proposed here the formal solution of Eq.~(\ref{E43}) can be expressed by
\begin{equation}\label{E44}
\tilde{F}_{M}(x, t) = \int_{-\infty}^{\infty} \tilde{p}_{M}(y, t) e^{- y \mathcal{\hat{O}}_{M}} \tilde{f}(x) dy,
\end{equation}
where the kernel $\tilde{p}_{M}(y, t)$ has to be adapted to a precise form of the operator $\hat{O}_{M}$. The Eq. (\ref{E43}) encompasses a large class of Fokker-Planck type operators, see \cite{KGorska12b} for related considerations. Furthermore the method can be shown to be applicable to the solution of problems where fractional evolution differential equations occur as in the case of anomalous diffusion \cite{KGorska12b} and relativistic quantum mechanics \cite{DBabusci11b, KKowalski11}.

\section{Conclusions}\label{Conc}

In this paper we have shown that the formalism of evolution equation and of the associated integral transforms is a very efficient tool to deal with evolution problems involving generalizations of the heat equations through the introduction of higher-order derivatives. We have seen how the formalism is capable of including \textit{popular} transforms like Gauss-Weierstrass and Airy via the so-called signed L\'{e}vy stable and generalized Airy Ai functions. 

The key result of the paper is the construction of a new technique of solving the HOHTE which furnishes the long-time behavior of Eq.~(\ref{E3}). We have also shown that our technique reconstructs the Hermite-Kamp\'{e} de F\'{e}ri\'{e}t polynomials being the formal solution of Eq. (\ref{E1}) with the initial condition $f(x)~=~x^{n}$. For the initial condition given by the L\'{e}vy signed function and the generalized Airy Ai function we observe the \textit{scaling} character of the time evolutions which are the natural extension of Glaisher-type relations. The next interesting result is related to the existence the border time in which the time evolution calculated for the Cauchy distribution begins to possess the negative parts. 

Most of the formalism developed in the paper can be applied to non-standard forms of evolution equations which are encountered in physical problems concerning anomalous diffusion and quantum mechanical relativistic effects. Regarding the first point, we note that many problems concerning the anomalous transport (in particular sub-diffusive) can be treated using HOHTE with not necessarily integer derivatives. Fractional transport is within the capabilities of the present formalism, which potentially offers possibility of treating in a unified way different phenomena occurring in economics \cite{RNMantegna95, BDMalamud04}, population mobility \cite{DBrockmann06, NHHumphries10}, infectious disease propagation \cite{HNymeyer98}, metastatic cancer spread \cite{SEMaier10, MLYizraeli11} \textit{etc.}

Regarding the more genuine physical aspects, we believe that the methods we have explored may certainly help illuminating old problems in relativistic quantum mechanics like some of its non-local feature, occurring \textit{e.g.} in the analysis of the relativistic Schr\"{o}dinger equation (see Refs. \cite{DBabusci11b, KKowalski11}, where some aspects of the underlying problems have started to be explored.) Before closing the paper we want to mention the possibility of looking at old problems with fresh eyes. The present formalism may yield unique tools to merge two aspects of anomalous diffusion and non-local quantum mechanics through the emergent L\'{e}vy generators \cite{PGarbaczewski13}, non-local in nature, which are naturally suited to provide a bridge between anomalous transport and pseudo-differential evolution in semi-relativistic quantum mechanics. It will be discussed in a forthcoming investigation.

\section{Acknowledgements}
We thank Prof. G. H. E. Duchamp and Dr. {\L}. Bratek for important discussions.

The authors have been supported by the Program PHYSCOMB no. ANR-08-BLAN-0243-2 of Agence Nationale de la Recherche (Paris, France) and by the PHC Polonium, Campus France, project no. 288372A.


\end{document}